# Inter-ELM pedestal evolution on MAST and impact of Resonant Magnetic Perturbations

R. Scannell, A. Kirk, I. T. Chapman

*EURATOM/CCFE Fusion Association, Culham Science Centre, Abingdon, Oxon OX14 3DB, UK*

E-mail: rory.scannell@ccfe.ac.uk

**Abstract** The peak pressure gradient in the pedestal ($dP_e/d\psi$) on MAST varies little between ELMs, although it varies between discharges due factors such as gas fuelling and plasma current. The pressure pedestal width in flux space on the high field side (HFS), both during the inter-ELM period and amongst different plasma discharges is consistent with a scaling of $\Delta_{pe} \approx 0.07\ \beta_\theta^{1/2}$. In flux space very similar $dP_e/d\psi$ and $\Delta_{pe}$ are observed on the HFS and low field side (LFS) in single null configuration. This symmetry is broken by the application of resonant magnetic perturbations (RMPs). During ELM mitigation by application of RMPs changes in the edge transport barrier position and width are observed. These changes are dependent on the intensity of the RMP and on the toroidal location with respect to the RMP phase. An outward displacement of up to 30mm and increase in the edge pedestal width of up to 50% with respect to the coils case off are observed. Increased particle transport causes a decrease in $n_{e,ped}$, and hence $P_{e,ped}$, as is observed on other devices. The combination of an increase in $\Delta_{pe}$ on the LFS and decrease in $P_{e,ped}$ results in significantly reduced LFS $dP_e/d\psi$ when these perturbations are applied to the plasma edge. A decrease of $dP_e/d\psi$ on the HFS is also observed due to RMP, however, this decrease is caused solely by the $P_{e,ped}$ decrease whilst no expansion of $\Delta_{pe}$ on the HFS is observed.

## 1. Introduction

Significant research effort has focused on using the pedestal height and width observed on current devices to predict pedestal and hence core plasma performance on future devices [1][2] [3] [4]. Improved pedestal performance implies increased $P_{e,ped}$, which in turn can increase particle and heat flux during ELMs. Recent experiments have used resonant magnetic perturbations (RMPs) applied to the plasma edge to mitigate and suppress ELMs [5][6][7] and increasingly there is a focus on the impact of RMPs on the pedestal [8][9]. The aim of the work presented here is to examine the implication of RMPs on the MAST pedestal behavior, particularly looking at the impact on pedestal width and sustainable pressure gradients in the edge transport barrier. This paper initially focuses on observations of the pedestal in a number of ELMing plasmas with no RMP and subsequently observations in plasmas with various configurations and intensities of RMP.

A typical MAST type I ELMing discharge has been analysed in some detail[10]. Stability analysis of this case has shown that this plasma becomes unstable to finite-n ballooning modes at the end of the inter-ELM period, triggering the ELM. However, it is the decreasing stability limit due to the region of high pressure gradient moving inwards, rather than increasing magnitude of the pressure gradient that triggers the ELM. The high gradient region is unstable to $n=\infty$ ballooning modes throughout the inter-ELM period. Gyrokinetic analysis [10] has shown that this region which is unstable to $n=\infty$ ballooning modes corresponds well to the region unstable to kinetic ballooning modes (KBMs), making these modes the likely candidate for limiting the pressure gradient within the edge transport barrier. In the first part of this paper, we compare the inter-ELM pedestal evolution of the standard MAST discharge with discharges with varying collisionality and magnetic configuration.

A number of tokamaks have demonstrated ELM mitigation due to RMP in the form of a large increase in ELM frequency [6] [11]. This increased ELM frequency causes a decrease in energy per ELM and hence decreased power loads on plasma facing components. ELM suppression due to RMP, complete removal of type I ELMs, has been achieved in DIII-D [5] and mitigation to very small ELMs on ASDEX Upgrade [7]. Although suppression is typically the ultimate goal, it may be on future devices that a regime including ELM mitigation is used either due to the inability to completely suppress ELMs or because some ELMs are required for impurity control. On MAST application of RMPs causes ELM mitigation, though complete suppression has not yet been observed[6]. The impact of n=6

RMP on the pedestal of a single null plasma is the focus of section 3 of this paper, detailing the impact on the radial profiles, evolution of pedestal parameters between ELMs and high field side (HFS) low field side (LFS) symmetry. For comparison purposes all of the RMP pulses are paired with equivalent pulses which have no applied RMP, but are otherwise identical. The final part of the paper examines the impact of two phases of applied RMP on a double null (DND) plasma, where phase is the toroidal modulation of the RMP with respect to the fixed toroidal location of Thomson scattering (TS) measurement.

The data presented here are predominantly taken from the MAST TS system [12], which measures with better than 10mm resolution across the full plasma radius. In all cases the data are deconvolved following the procedure outlined in [13] to remove the impact of the instrument function. Deconvolution does not impact on the conclusions from the results discussed in this paper. In this deconvolution procedure, the fits are performed to $T_e(r)$ and $n_e(r)$ in real space, where the instrument function can be removed, the results are then mapped to flux space where the pressure profile is then fitted. Hence throughout this paper the density and temperature pedestal widths ($\Delta_{Te}$ and $\Delta_{ne}$) are referred to in real space and the pressure pedestal width ($\Delta_{pe,\psi}$) is referred to in flux space. The same trends of pedestal width expansion with $\beta_{poloidal}$ and width expansion due to application of ELM coils are observed in the profiles in both real and flux space. Fitting to the edge radial profiles is performed using the modified tanh function [14] and the pedestal width, height and position obtained from these fits are used throughout this paper.

## 2. Edge evolution in the inter-ELM period

In the period between typical MAST type I ELMs [10] an increase in $P_{e,ped}$, an increase in $\Delta_{pe,\psi}$ and inward movement of the transport barrier position are observed. The combination of increasing $P_{e,ped}$ and expanding $\Delta_{pe,\psi}$ results in only a small change in peak pressure gradient between ELMs. The inter-ELM period in these discharges, which have $T_{e,ped}$ of ~150eV, is compared with that of a high temperature pedestal (up to ~300eV) or low collisionality discharge. The low collisionality discharge, achieved with lower gas fuelling and a modified current ramp up, has a $\nu^*_e$~0.35, a $q_{95}$ of ~5.7, elongation of 1.9, line average density $<n_e>=2.9x10^{19}m^{-3}$, a plasma current of 800kA and $B_T$ on axis of 0.51T. The 150eV pedestal discharges have $\nu^*_e$~1.1, $q_{95}$~4.8, elongation of 1.7, line average density $<n_e>=3.5x10^{19}m^{-3}$, a plasma current of 850kA and a $B_T$ on axis of 0.47T. Both discharges had similar heating, two neutral beams with a total injected power of 3.4MW. In the low collisionality discharge the temperature pedestal height as well as the density pedestal height increases between ELMs. A comparison of the pedestal pressure parameter evolution for the two discharges after the ELM are shown in figure 1. During the inter-ELM period for both discharges the position of the edge transport barrier moves inwards and $P_{e,ped}$ and $\Delta_{pe,\psi}$ increase. The pressure gradient profiles, measured in the final 25% of the inter-ELM period are shown in figure 1(d). The peak $dP_e/d\psi$ obtained in the low collisionality discharge is significantly higher and occurs further inside the plasma. Gyrokinetic analysis of this low collisionality discharge shows that unlike in the high collisionality case[15] the pedestal region is stable to KBMs due the high bootstrap current which causes low magnetic shear [16].

The EPED [17] [18] model has been used to predict the pedestal parameters on a number of different devices [19]. In this model, the pedestal gradient is limited by KBMs, and hence pressure pedestal width scales with pedestal poloidal beta as $\Delta_{pe,\psi}\propto\beta_{poloidal,ped}^{1/2}$. Pressure pedestal height is separately predicted from peeling-ballooning mode stability and combining these two constraints gives the pedestal operating point. Further to this there has been a number of empirical results showing pedestal width scaling approximately with $\beta_{poloidal}^{x}$, where x~1/2 [3][4][20]. Figure 2 shows the pedestal top and pedestal width evolutions for four different MAST discharges: 400kA and 650kA single null discharges and the two double null discharges discussed earlier in this section. Averaged results of the pedestal parameter evolution at four bins (0-25%, 25-50%, 50-75% and 75-100%) in the inter-ELM period are shown for each discharge. Different symbols represent the different discharges and the size of each symbol represents increasing time within the inter-ELM period. For all discharges except the low $\nu^*$ DND discharge pedestal evolution throughout the inter-ELM period implies an increasing $n_{e,ped}$ at constant $T_{e,ped}$. Similarly increasing plasma current from the 400kA to the 600kA in SND discharges causes an increase of edge pressure through $n_{e,ped}$. The period 75-100% of the inter-ELM period in the 400kA discharge has similar pedestal profiles to the period 25-50% in the 600kA discharge.

For the four discharges shown in figure 2 the high field side pedestal width scaling with $\beta_{poloidal}$ has been examined. All discharges have an injected neutral beam power of ~3.4MW. The discharges have a scaling of pedestal width in the range $\Delta_{pe,\psi} = 0.06\text{-}0.08\sqrt{\beta_{pol}}$, where the pedestal width is measured in normalized flux space. This scaling is valid both within the inter-ELM period for any given discharge and comparing the pedestal widths at the end of the inter-ELM period for different plasma discharges. The results also show that this scaling holds across magnetic configurations, that is for both single and double null plasmas with their very different geometries. The scaling also holds whether the variation in $\beta_{poloidal}$ is caused by variation in $P_{e,ped}$ at approximately constant $I_p$ as in the high and low $\nu^*$ DND plasmas or variation in $P_{e,ped}$ as well as $I_p$, as in 400kA and 600kA discharges. For this analysis the magnitude of poloidal magnetic field at the plasma circumference is estimated as $B_{pol}=\mu_0 I_p/L_p$ [20] where $L_p$ is the poloidal circumference of the last closed flux surface.

The pressure pedestal widths $\Delta_{pe,\psi}$ and peak gradients $dP_e/d\psi$ on the high field side (HFS) and low field side (LFS) behave differently depending on the plasma configuration, as shown in figure 3. As would be expected, for both SND and DND plasmas the pedestal top value, $P_{e,ped}$, observed on the HFS and LFS are identical. In SND plasmas, very similar $\Delta_{pe,\psi}$ and hence similar $dP_e/d\psi$ are observed on the HFS and LFS. By contrast in DND plasmas broader $\Delta_{pe,\psi}$ and lower $dP_e/d\psi$ are observed on the LFS of plasma in flux space. There is typically a flux expansion of ~2 from HFS to LFS, which means that in real space there is good agreement in pedestal width LFS to HFS in the DND case and similarly for the pressure gradient as observed in real space. Similar results have been reported previously on MAST [3] obtained using different Thomson scattering systems (a Ruby TS system and edge Thomson system) giving a very high degree of confidence in this observation. An implication of different pedestal widths at different poloidal locations in these DND plasmas is that devices where the pressure profile is measured away from the LFS midplane and then mapped using normalized flux may not correspond to actual midplane profiles as is implicitly assumed in the analysis.

## 3. Application of n=6 RMP field to SND discharges

*3.1 Gradient change on application of n=6 RMP field*

Figure 4 shows $P_{e,ped}$, peak LFS $dP_e/d\psi$, plasma energy and $D_\alpha$ emission as a function of time for plasma discharges with and without applied RMP field. Similar, but not identical, trends are observed on the HFS as discussed later in this paper. In figure 4 the discharge with the applied RMP field shows a much higher ELM frequency. A decrease in both $P_{e,ped}$ and $dP_e/d\psi$ in the pedestal are observed when the ELM coils are applied. Only points from ELMs in the last 80% of the inter-ELM period are shown, as the gradient observed in the range 0-20% is highly variable. The scatter in the $P_{e,ped}$ and $dP_e/d\psi$ shown is due to uncertainty in the measurement and timing of the measurement in the inter-ELM period. During the inter-ELM period $P_{e,ped}$ varies by a factor of two and $dP_e/d\psi$ by ~15-20%. However the change in $dP_e/d\psi$ upon application of the RMP is much larger than the measurement scatter. Cases are also observed where the RMP field is applied before the L-H transition, these cases show a lower $dP_e/d\psi$ throughout the discharge than equivalent discharges with no RMP.

*3.2 Radial Profiles on application of n=6 RMP field*

The radial profiles from two sets of discharges are compared: an ‘RMP on’ dataset with applied n=6 perturbations and an ‘RMP off’ dataset with no applied field. The RMP on dataset contains 69 profiles taken during mitigated H-mode periods of 30ms to 120ms during three MAST discharges #27204, #27811 and #28002. The RMP off data set contains 75 profiles during three MAST discharges #27205, #27315 and #28158. Other than the current in the RMP coils, these discharges are identical. These discharges had a plasma current of 620kA, a $q_{95}$ of 2.6 and a pedestal top collisionality of 1.2 and are all in the lower single null (SND) configuration. Both datasets have moderate gas fuelling rates (the implications of different fuelling rates will be discussed later in the paper). The ELM mitigation observed in the RMP on dataset is that there is an increase in ELM frequency from approximately 60Hz to 200Hz.

For these RMP on and RMP off datasets the radial profiles in the last 50% of the inter-ELM period have been averaged (figure 5). A reduction in $n_{e,ped}$ is observed due to the RMP field caused by increased particle transport, or some other mechanism, however the two sets of shots have similar $T_e$ profiles. It can be seen that the plasma edge is moved outwards by ~30mm, equivalent to ~6% r/a, on turning on the coils. The observed LFS $\Delta_{ne}$ in the 'RMP on' shot is increased by 50% with respect to the 'RMP off' shot as the plasma edge is moved outwards, an increase in $\Delta_{Te}$ is also observed. In general decreasing $\beta$ should cause a decrease in $\Delta_{\psi}$ as per the scaling, hence the increase in $\Delta$ (in real space here, but also in flux space as shown later) appears to be a result of RMPs. The HFS of the plasma experiences the density pump out but shows no evidence of change in position or pedestal width due to the application of the RMP field.

The outward movement of the plasma is a toroidally asymmetric perturbation[21] but also includes a plasma control system response due to a drop in $\beta$. This system response can be removed by looking at the relative edge radial displacement at two toroidal locations. Figure 6(a) shows the relative radial position of the ETB from mtanh fits to TS data and the $D_\alpha$ peak measurement with and without applied RMPs. The $D_\alpha$ camera images the outboard midplane of the MAST plasma with high resolution, for a full description see[22]. The $D_\alpha$ camera lines of sight go through the vertical midplane, while the TS diagnostic measures 15mm above the vertical midplane. In these single null discharges, the plasma center is at z=-200mm and the LCFS of the plasma is steeply angled at the vertical midplane. For this reason the ETB location seen by the TS diagnostic is ~10mm further inside that seen by the $D_\alpha$ camera in the 'RMP off' case. The toroidal measurement locations of TS and linear $D_\alpha$ camera are indicated on figure 6(b). The relative radial displacements at the two toroidal locations are compared with results of the vacuum field line tracing code ERGOS [23] in figure 6 (b). These plots show field lines traced for 200 toroidal turns as a function of radial location and toroidal angle. The line colours indicate the minimum normalised flux that the field experiences during its trajectory. In this case the TS measures at close to the maximum edge displacement due to the n=6 RMP and the $D_\alpha$ close to the minimum. Using this $\psi$ as a function of radius and toroidal angle from ERGOS the radial displacement of the ETB at the toroidal locations of the $D_\alpha$ and TS diagnostics is calculated to be ~18mm, which agrees well with the direct measurements in figure 6(a).

*3.3 Pedestal profile evolution during application of n=6 RMP fields*

Figure 7 shows the pedestal parameter evolution for the RMP on and RMP off datasets for data collected from the LFS of the plasma. The RMP on and off datasets are binned as a function of time through the inter-ELM period. Comparing the pedestal top values at the same time after the ELM, the $T_{e,ped}$ is similar while $n_{e,ped}$ and $P_{e,ped}$ show a reduction of approximately 10% with the RMP on. The density and temperature pedestal widths as measured in realspace at the same time after an ELM are observed to be significantly larger in the RMP on case, with the exception of the first bin after the ELM collapse. However, for up to 1ms after the ELM it is difficult to accurately measure pedestal parameters, so this bin is not particularly useful for comparison purposes. As well as expansion of $\Delta_{ne}$ and $\Delta_{Te}$ in realspace due to applied RMP, using EFIT mapping to normalised flux without knowledge of a toroidal perturbation, a similar expansion of $\Delta_{pe,\psi}$ is observed. A phenomenon such as island structures, reported to exist in the pedestal during RMP experiments on DIII-D [5], could lead to increased pedestal widths being observed.

A comparison of $dP_e/d\psi$ on the HFS and LFS of the plasma before and after application of the RMP field is shown in figure 8. With no applied field, similar pedestal widths and gradients are observed HFS and LFS, as is typical for all SND discharges on MAST. Upon application of the applied field there is a collapse in pressure gradient of 30-50% due to $n_{e,ped}$ decrease on both high and low field sides of the plasma. This collapse in gradient is larger on the low field side due to an increase in pedestal width, whilst no increase in pedestal width is observed on the high field side.

**4. Variation of pressure gradient with fuelling and RMP**

Discharges with different gas fuelling levels and RMP coil currents, with applied fields of both n=4 and n=6 have been examined and the results are summarised in figure 9. All discharges have the same injected neutral beam power of 3.4MW. With no additional gas fuelling coil currents of 1kA cause approximately a doubling in ELM frequency and currents of 1.4kA typically result in a back transition to L-mode. In shots with refuelling, currents of 1.4kA cause a similar increase in ELM frequency and no back transition, for more details see [24]. Hence in figure 9, a discharge without refuelling and 1kA of coil current is compared with discharges with refuelling and 1.4kA of coil current.

The $n_e$-$T_e$ diagram in figure 9 shows that for a plasma no refuelling during the applied RMP, 1kA of RMP current causes a reduction in $n_{e,ped}$ and a significant increase in $T_{e,ped}$. The increase in $T_{e,ped}$ is likely due to increased heating power per particle in the lower density plasma. In the refuelling cases with 1.4kA in the RMP coils there are larger decreases in $n_{e,ped}$ and little or no increase in $T_{e,ped}$. The lack of increase in $T_{e,ped}$ may be due to the increase in heating power per particle being outweighed by the increased recycling due to higher neutral pressure. A moderate amount of gas fuelling, as seen in figure 9(e), is optimal for typical MAST discharges and allows them to reach higher $P_{e,ped}$ and $dP_e/d\psi$ than discharges with no fuelling or very high fuelling. Application of RMP field due to 1.4kA of coil current to these moderately fuelled shots reduces $P_{e,ped}$ and $dP_e/d\psi$ to the same level as 1.0kA of coil current in the unfuelled discharges.

An outwards displacement of the plasma edge and increase in pedestal width are observed on the LFS of the plasma in all discharges examined in figure 9(b). Edge displacement and pedestal width expansion both increase with increasing RMP coil current, although the effects of fuelling cannot be separated out.

**5. Application of n=3 fields to Double Null Discharges**

The toroidal variation of edge displacement can be examined by applying different phases of the same perturbation and measuring at a fixed toroidal location. In this section edge profiles during application of n=3 error fields with phases of 0 and 60 degrees are compared. Ideally measurements at the minimum and maximum of the radial perturbation due to the RMP would be obtained, however TS measurements in this case are somewhere between.

For discussion of DND plasmas ELM coil parity must be mentioned. There are two rows of ELM coils one above and one below the midplane. Even parity means that coils in the same toroidal location in upper and lower rows have the same current, odd parity means that these coils have the opposite current. For the DND plasmas discussed here the coils are in even parity. For SND plasmas discussed previously only the lower row of coils are active, hence parity is not an issue.

Figure 10(a) shows the relative radial position of the ETB from mtanh fits to TS data and the $D_\alpha$ peak measurement for different applied RMPs. For DND plasmas the centroid of the plasma is at z~0mm. Hence the 15mm vertical separation between the $D_\alpha$ and TS lines of slight has little impact on the observed radial position of the LCFS by the two diagnostics. For no applied field the ETB position from the two diagnostics agrees well. Application of an n=3 0degree perturbation causes a toroidally local outward movement of the TS ETB with respect to the $D_\alpha$ ETB of ~10mm for an n=3 0degree RMP field. An outward movement of ~6mm is observed for a n=3 60degree field.

The relative radial displacements at the two toroidal locations can be compared with the predictions of the vacuum field line tracing code ERGOS in figures 10 (b) and (c). The predicted magnitude of the edge perturbation in this case is ~10mm which is much smaller than the SND case (figure 6) where perturbations of 30mm are predicted by ERGOS. The smaller offset in position between $D_\alpha$ and TS which is observed, agrees with the ERGOS prediction. The relative change in TS and $D_\alpha$ ETB in going from n=3 0 to 60degree applied

RMPs also agrees with the ERGOS prediction. However, the offset between the TS and $D_\alpha$ ETB for the coils on cases relative to the coils off cases is larger than that predicted by ERGOS. A further experiment is planned where the n=3 perturbation will be rotated in the view of the TS and $D_\alpha$ diagnostics, which will provide more detailed toroidal sampling of the RMP edge perturbation.

The LFS pedestal evolution in the inter-ELM for this DND case, shown in figure 11(a) and (b), is similar to SND discharges with ELM coils: $n_{e,ped}$ increases in line with the 'RMP off' $n_{e,ped}$ except that the increase is terminated by an ELM sooner and larger pedestal widths are observed with RMP on. The observed $\Delta_{ne}$ for n=3 0 degree RMP is marginally larger than those for the 60 degree phase shifted RMP. Although the width difference between the two phases of perturbation is within the measurement uncertainty, $\Delta_{ne}$ are clearly larger with RMP on than off.

Figure 11(c) shows the stability analysis using ELITE [25] [26] for an RMP off case and a n=3 0 degree applied field case for individual profiles taken in a 1ms time window before an ELM is triggered. Without applied RMP perturbations, the plasma is very close to the stability limit just before the ELM. With applied RMP perturbation the plasma is far from the stability limit, hence the subsequent triggering of an ELM must be explained by other means. ELMs triggered in this stable region during application of RMP have also been observed in SND plasmas [27], where it is proposed that perturbations near the x-point are the mechanism for degrading the peeling-ballooning stability boundary.

**6. Summary**

The first part of this paper examines discharges without RMP. In these discharges, the pressure gradient in the pedestal does not change significantly during the final 75% of the inter-ELM period. For a given discharge, this approximately constant $dP_e/d\psi$ during the inter-ELM period occurs because $\Delta_{pe,\psi}$ increases with $\beta_\theta$ to compensate for $P_{e,ped}$ increase between ELMs. The same pedestal width scaling observed during the inter-ELM period of a given discharge is also observed between discharges at different $I_p$ and for different magnetic configurations. In single null plasmas the same $\Delta_{pe,\psi}$ and $dP_e/d\psi$ are observed at the high and low field sides of the plasma. However, in double null plasmas $\Delta_{pe,\psi}$ is larger and hence $dP_e/d\psi$ is lower on low field side of the plasma. This has implications for other machines where measurements of the transport barrier are mapped to the low field side mid-plane.

Applying an RMP field to the plasma causes a particle 'pump out' resulting in a significant decrease in pressure gradients. On the low field side of the plasma the RMP also causes toroidally local outward displacements of the plasma edge and toroidally local increases in pedestal width. The pedestal width on the high field side of the plasma is unaffected by the perturbation which is observed only at the low field side. This results in poloidally asymmetric pressure gradients in SND discharges during RMPs. In the case of an n=6 RMP applied to an SND plasma a reduction of $dP_e/d\psi$ on the high field side of ~30% (decrease in $P_{e,ped}$, constant $\Delta_{pe,\psi}$) is observed as against ~50% on the low field side (decrease in $P_{e,ped}$, increase $\Delta_{pe,\psi}$). This strongly reduced pressure gradient would imply greater stability to MHD, but is observed to exist with increasing ELM frequency. Pedestal evolutions in the inter-ELM period are similar with and without the applied perturbation with the difference that, when there is an applied perturbation, the evolution is terminated by an ELM much sooner.

A second case examined was the application of n=3 field at toroidal angles of 0 and 60 degrees to a DND plasma. Much smaller LCFS perturbations are predicted by ERGOS for this case, ~10mm as opposed to ~30mm for the n=6 case. A smaller radial perturbation was observed between the TS and $D_\alpha$ diagnostics in line with the prediction. An expansion in pedestal width was also observed. Since the Thomson scattering system does not measure at the toroidal minimum of the RMP field it is not yet known if $\Delta_{pe,\psi}$ does not expand at this point and hence if there is a toroidally local higher $dP_e/d\psi$ at these RMP minima. Further experiments investigating the toroidal variation of the pressure profile during RMP are planned.

**Acknowledgements**

This work was funded by the RCUK Energy Programme under the grant EP/I501045 and the European Communities under the contract of Association between Euratom and CCFE. The views expressed herein do not necessarily reflect those of the European Commission.

**Figures:**

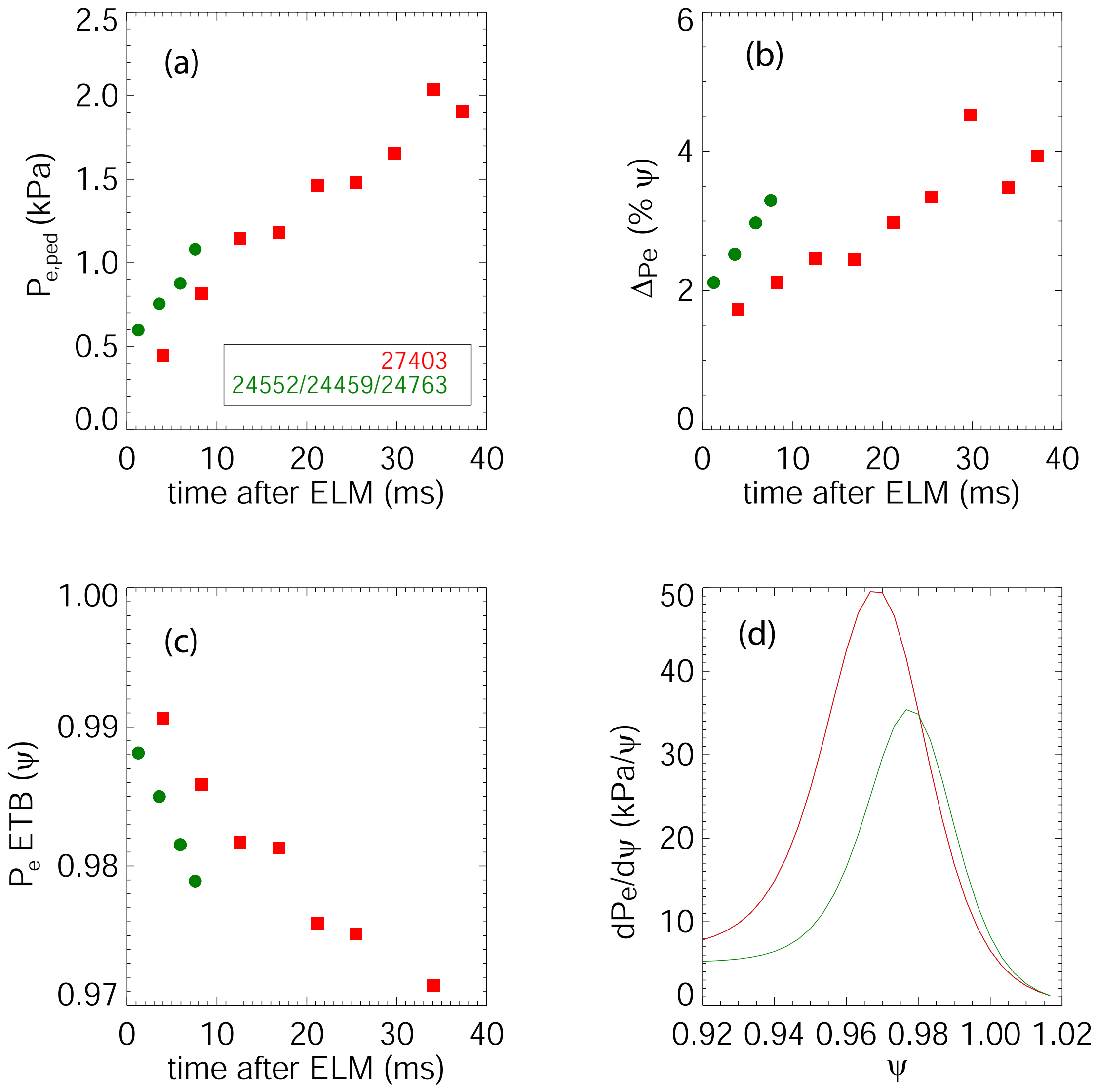

**Figure 1 - Pressure pedestal evolution between ELMs for a $T_{e,ped}$=150eV MAST H-mode discharge (green circles) and for a high temperature pedestal discharge (red squares) (a) pressure pedestal height (b) pedestal width (c) edge transport barrier position (d) pressure gradient profiles for the edge region in the final 25% of the inter-ELM period.**

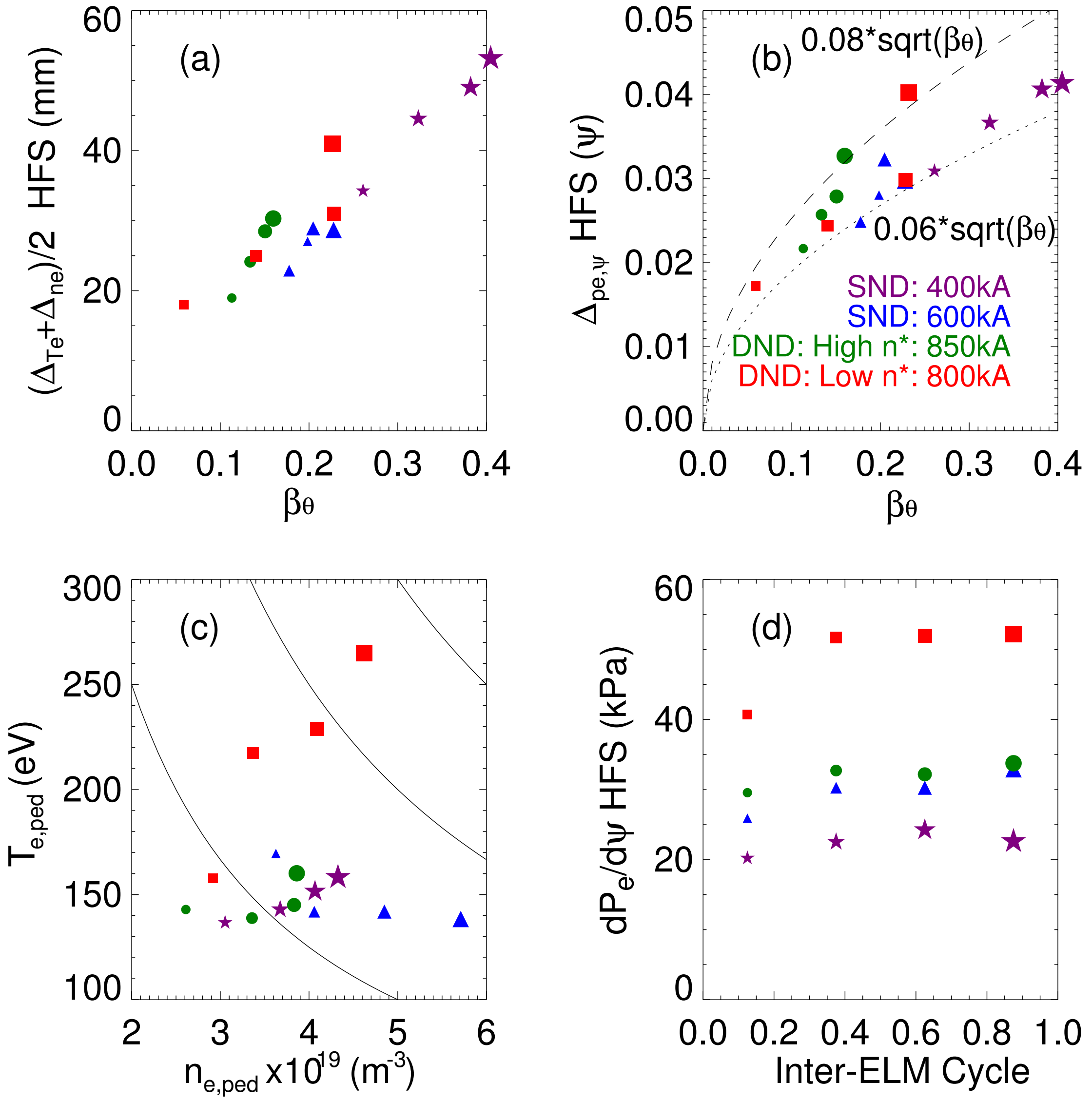

**Figure 2 - Four MAST discharges shown each represented by a different symbol. Each discharge is evaluated at four times during the inter-ELM period. Increasing time in the inter-ELM period is represented by increasing symbol size (a) evolution of average of density and temperature pedestal widths in real space (b) pressure pedestal width in flux versus poloidal beta (c) pedestal top electron temperature and density (d) pressure gradient evolution versus time with in the inter-ELM period**

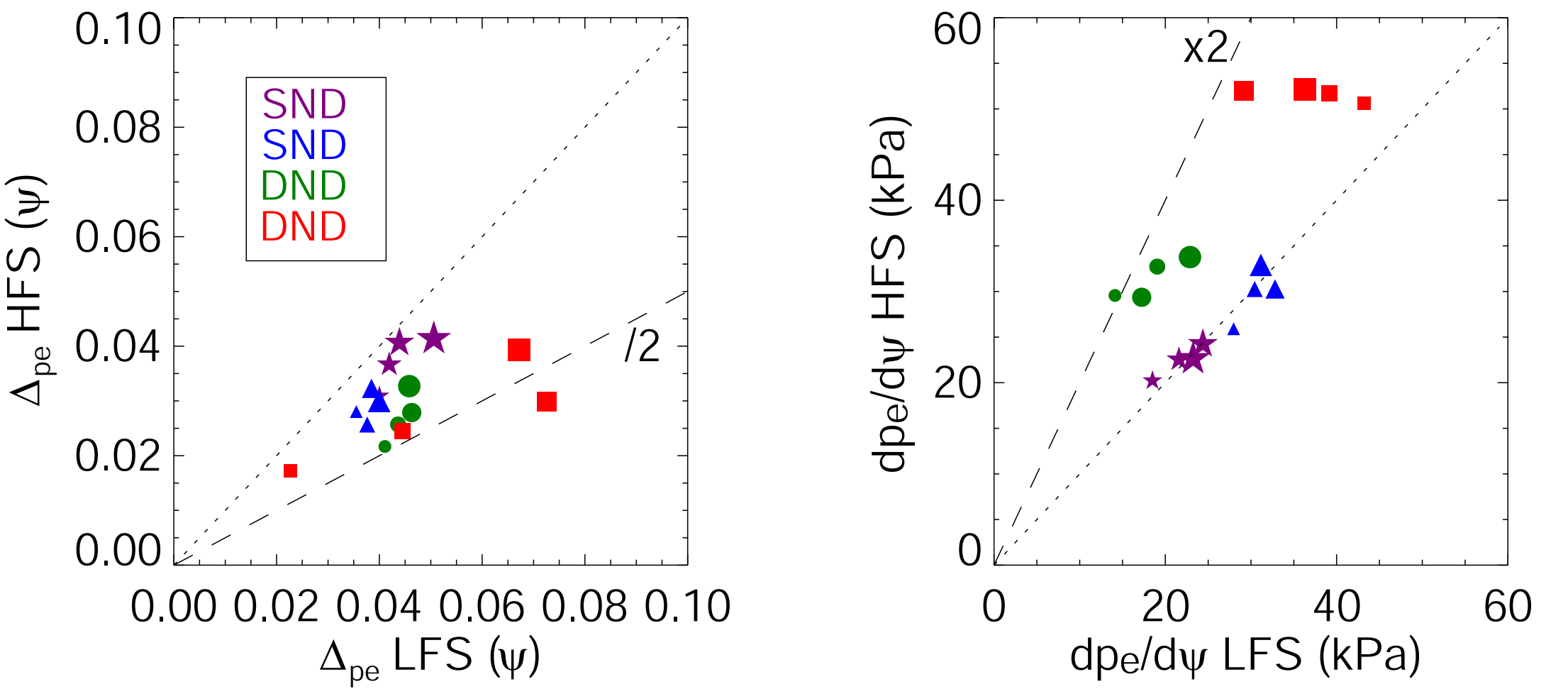

**Figure 3 –Comparison of the pedestal widths and pressure gradients on the high and low field sides of the plasma in single null (SND) and double null (DND) configurations for the same four discharges as in figure 2. Symbol size represents time during the inter-ELM period of the discharge.**

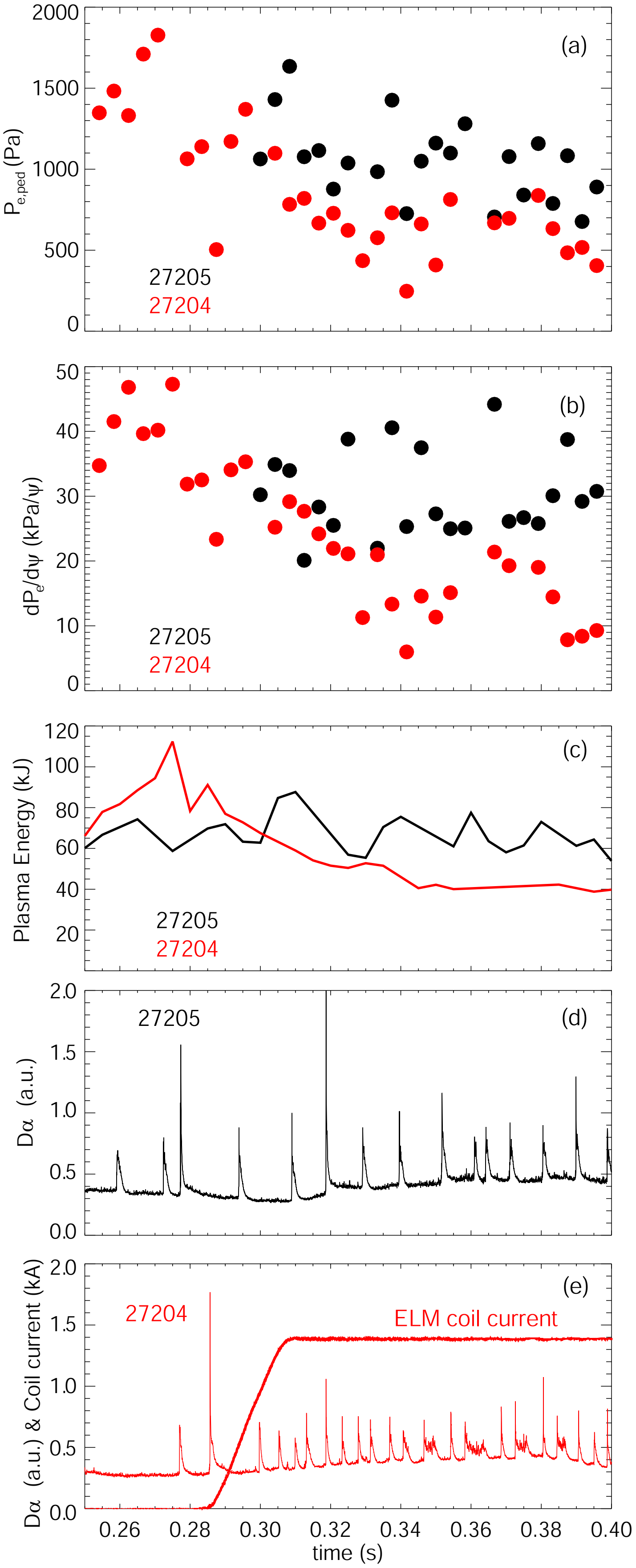

**Figure 4 – (a) Pressure pedestal height and (b) peak pressure gradient evolution for identical discharges with and without application of 1.4kA of RMP coil current applied in an n=6**

**configuration. (c) Total stored plasma energy from magnetic reconstruction (d) Dα emission for no RMP discharge (e) Dα emission and coil current for mitigated discharge.**

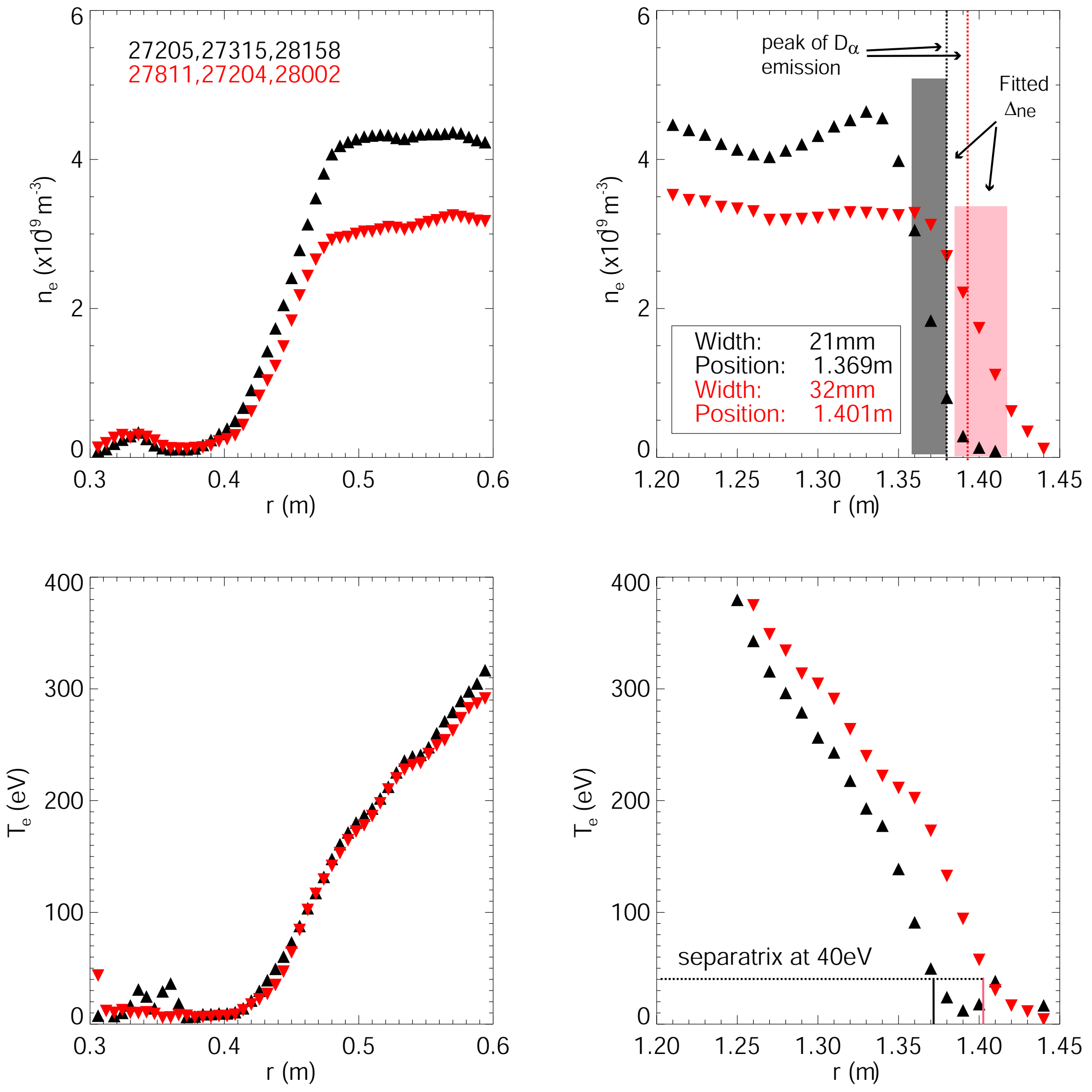

**Figure 5 – Profiles of edge electron temperature and density for shots without RMP (black) and with applied n=6 RMP fields (red) at the high and low field sides of the plasma.**

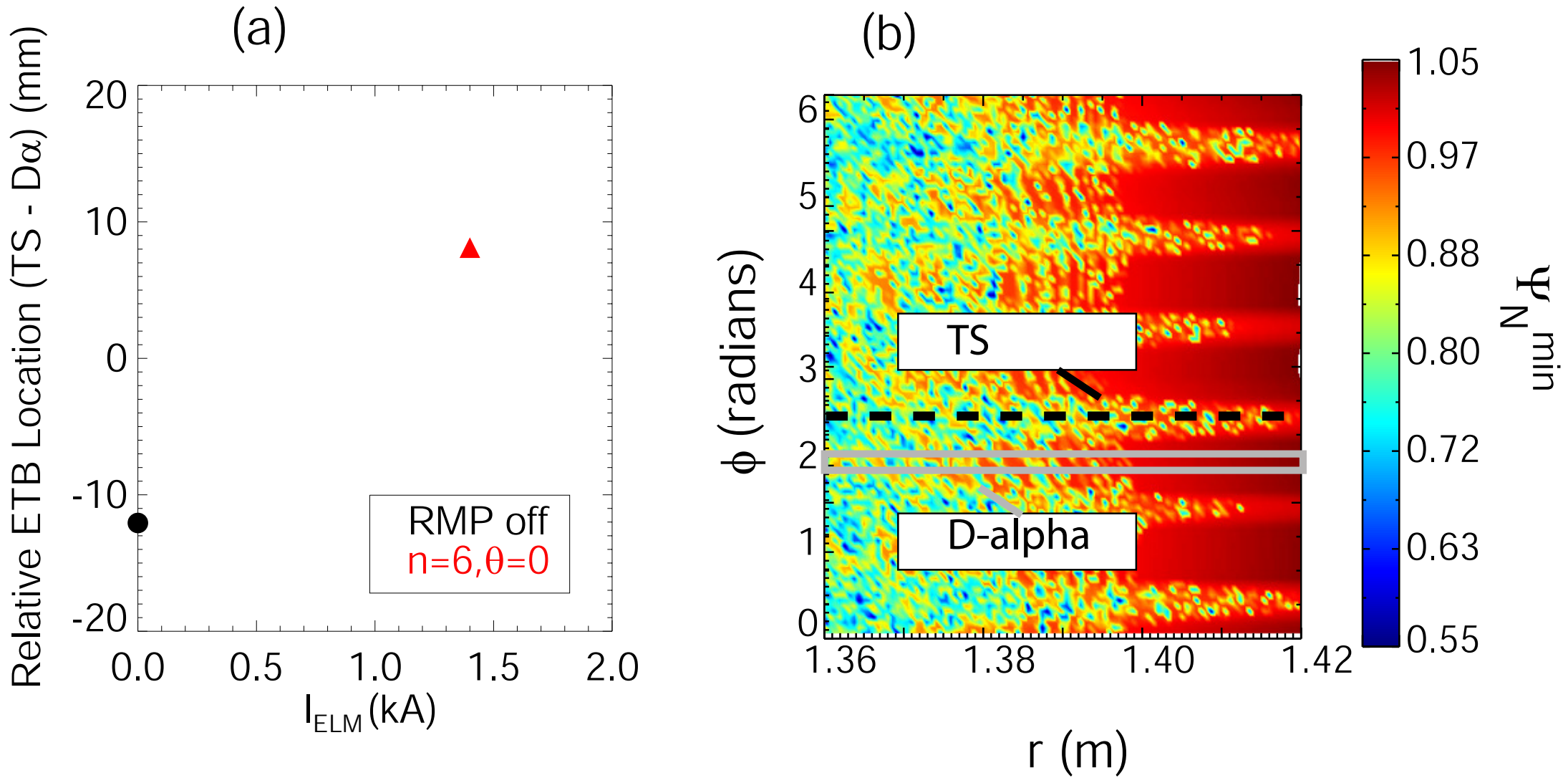

**Figure 6 – (a) Edge transport barrier position obtained from Thomson scattering profiles relative to the peak of $D_\alpha$ signal (b) ERGOS simulation of plasma with applied n=6 perturbations showing toroidal measurement locations of TS and $D_\alpha$ measurements.**

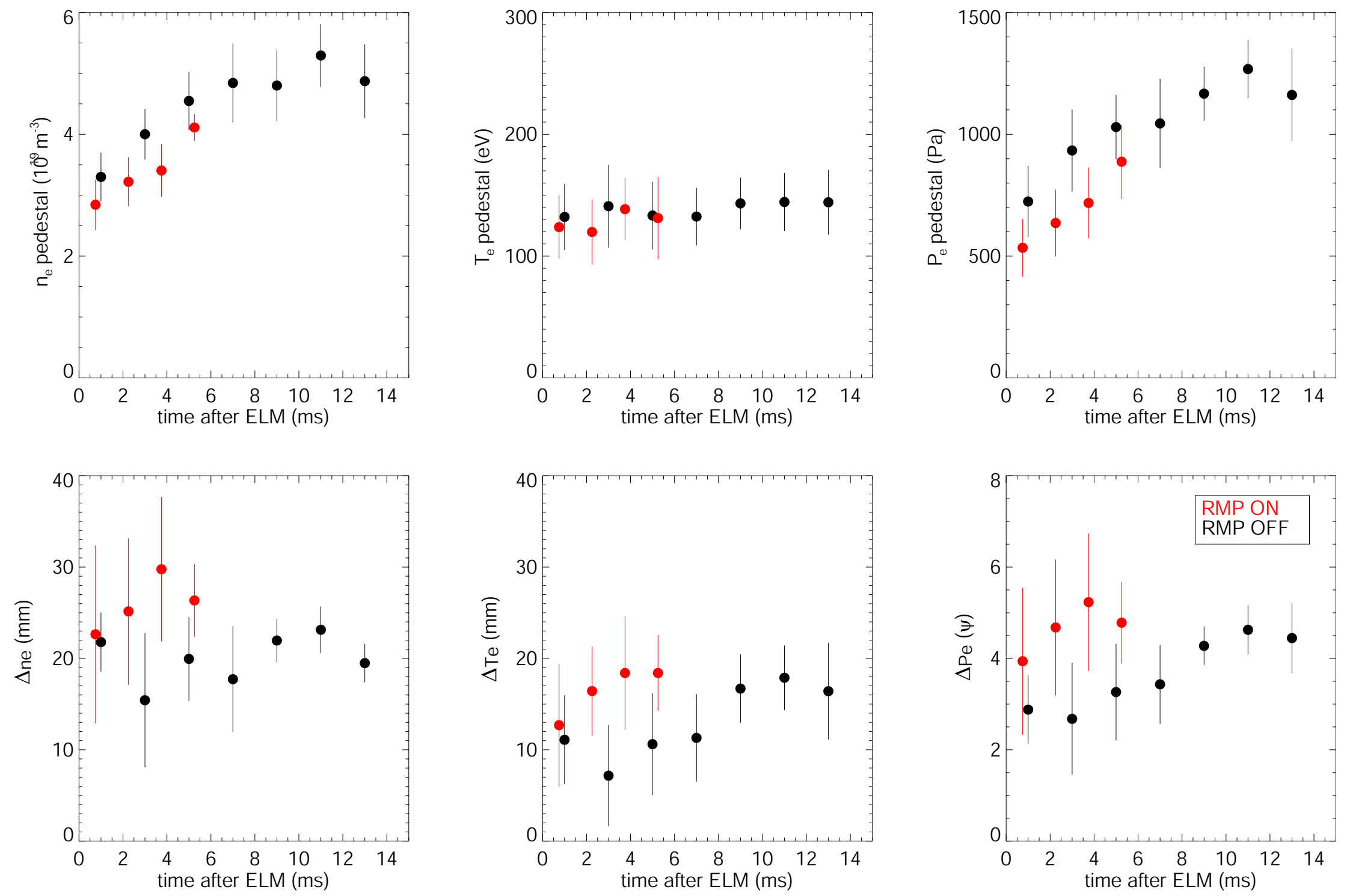

**Figure 7 – Evolution in the inter-ELM of pedestal parameters on low field side with RMP off and an RMP coil current of 1.4kA in an n=6 applied perturbation.**

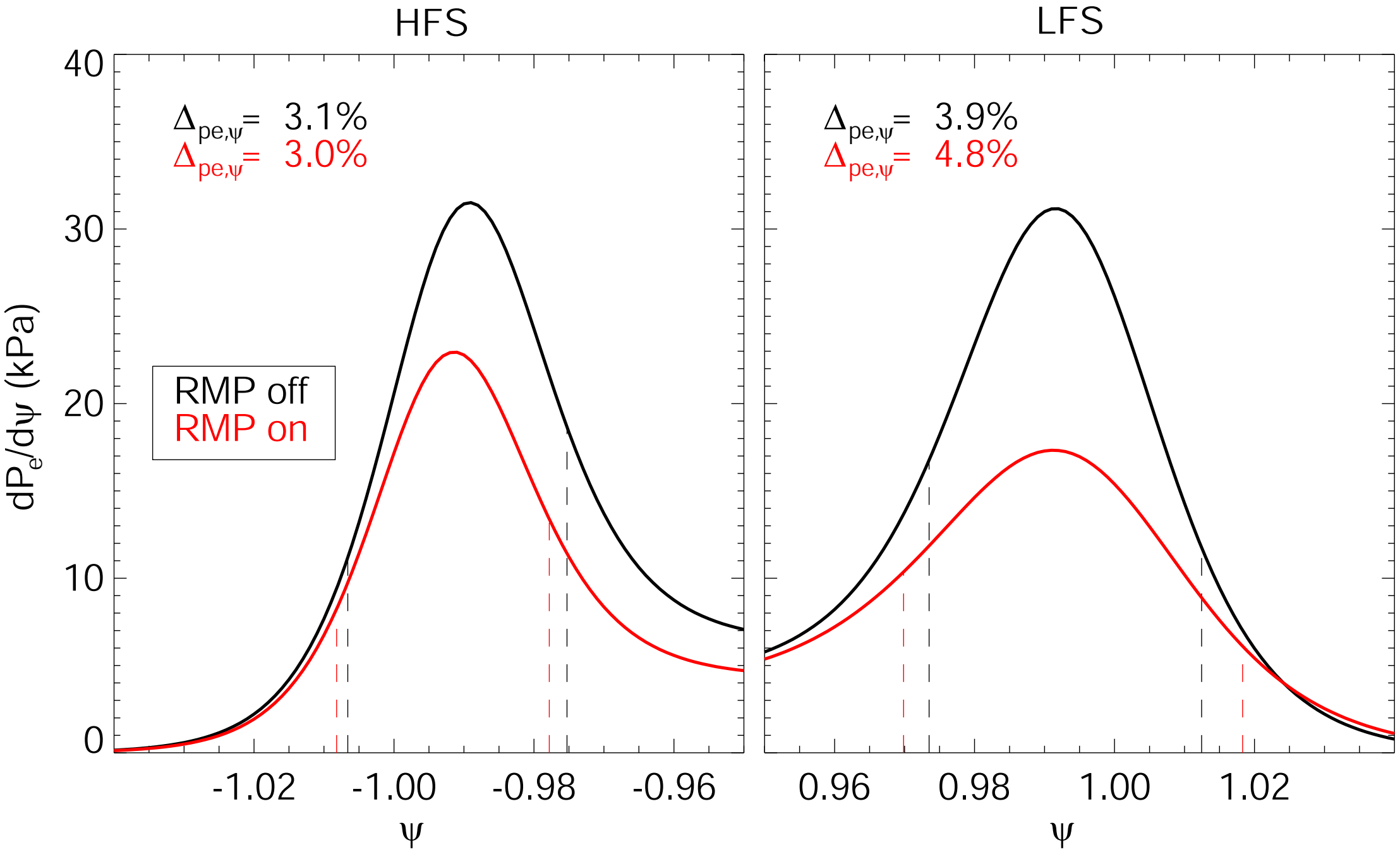

**Figure 8 – Comparison of pressure gradients within the pedestal with and without RMP coils as measured at the high and low field sides of the plasma. Vertical dashed lines indicate the measured pedestal widths.**

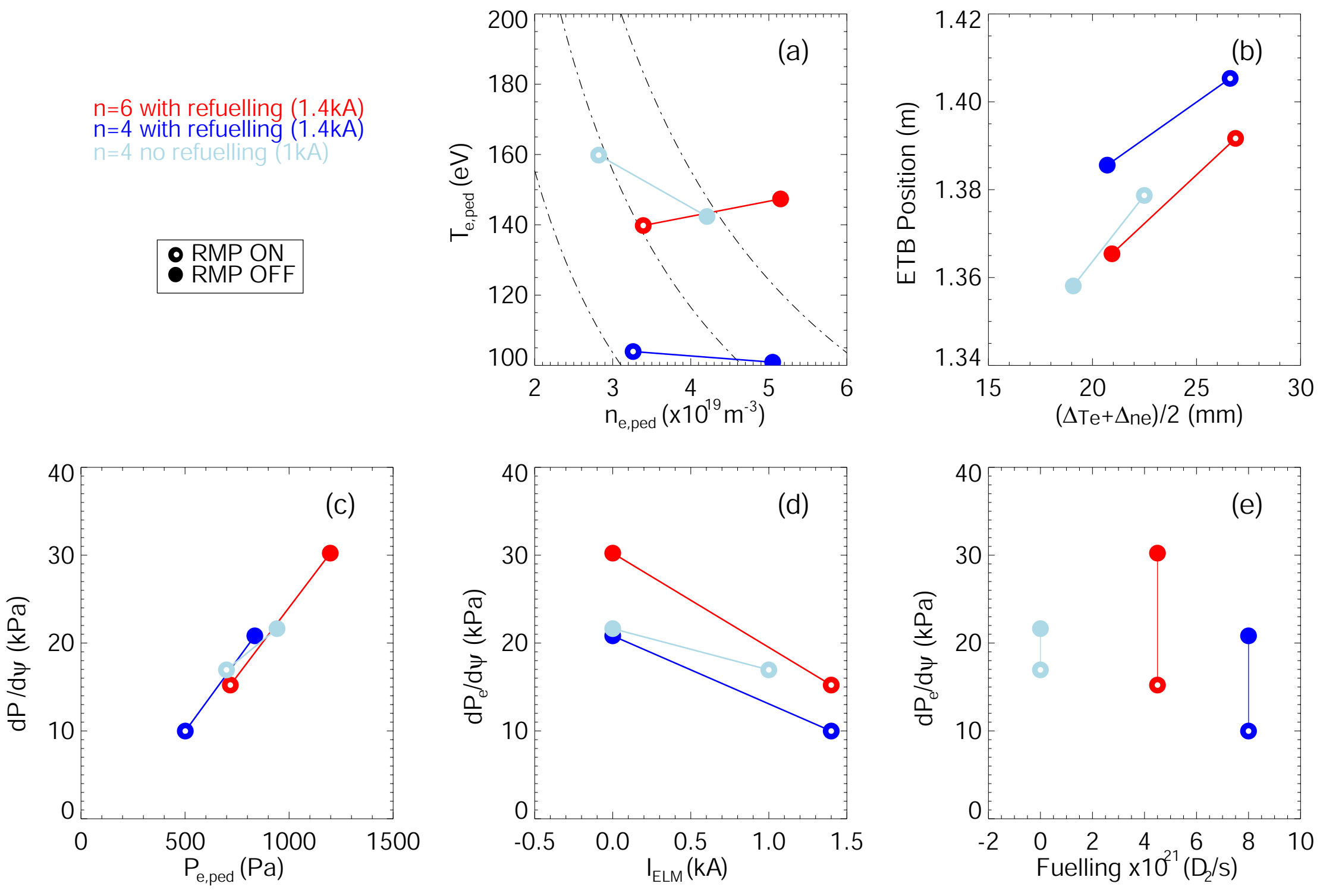

**Figure 9 – Impact of turning on ELM coils for three different MAST discharges. In all cases the filled symbol respresents the RMP off discharge and the hollow symbol the RMP on discharge. (a) Temperature and density pedestal. (b) edge position and pedestal width change (c) pressure gradient and pressure pedestal height (d) pressure gradient change with coil current (e) pressure gradient change with gas fuelling.**

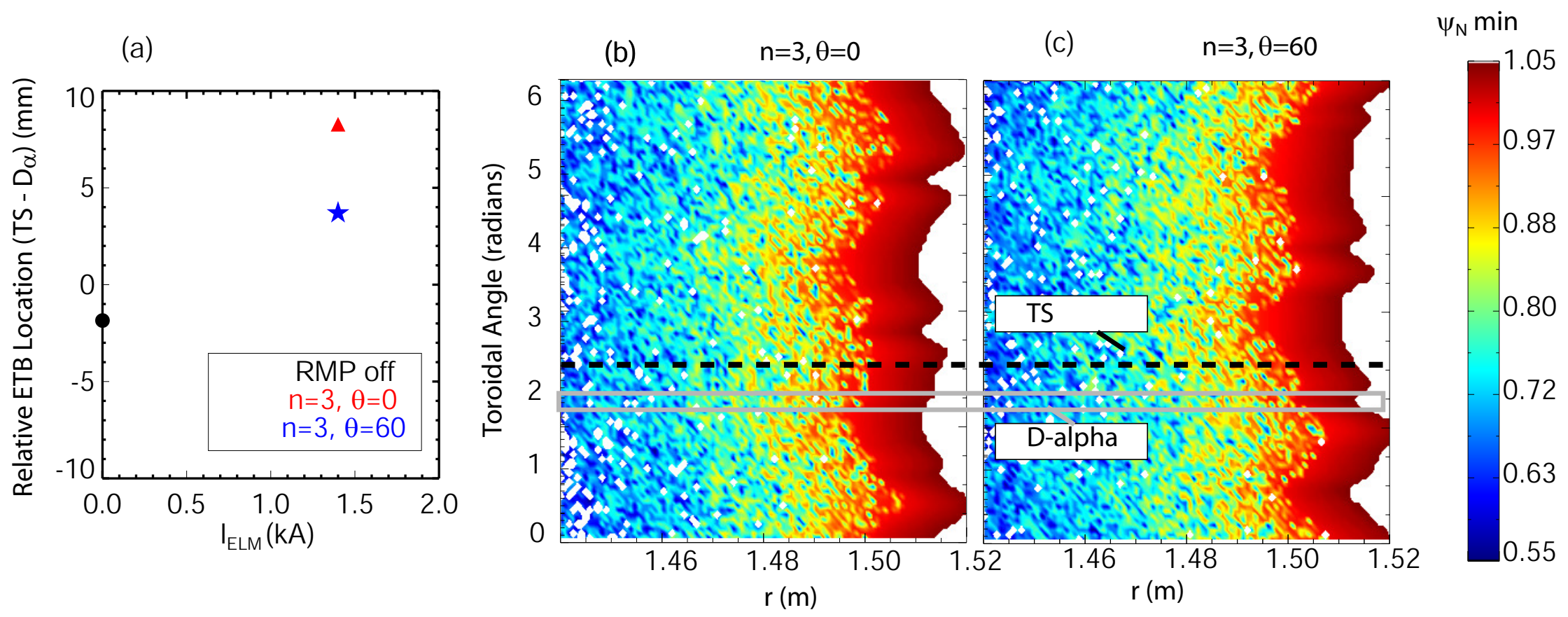

**Figure 10 – (a) change in position of edge transport barrier as measured by Thomson Scattering for application of n=3 error fields with a phase of 0 degrees and 60 degrees. (b) ERGOS simulation for case of n=3, 0 degrees (c) ERGOS simulation for the case of n=3, 60 degrees**

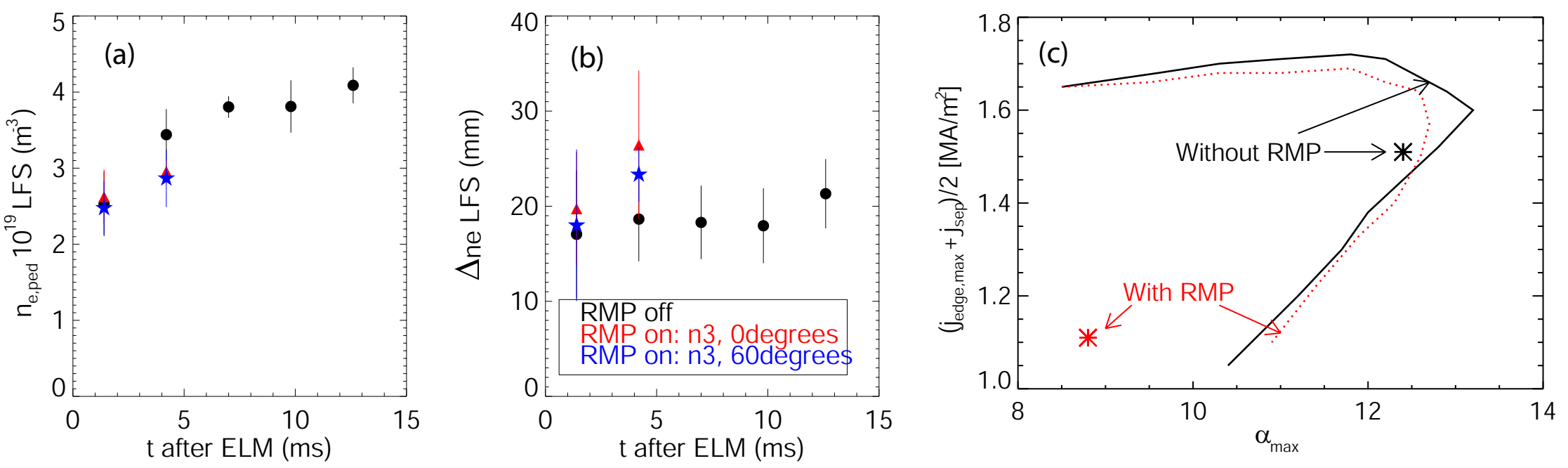

**Figure 11 – (a,b) Pedestal density and width evolution in the inter ELM for different phases of error fields. (c) edge stability to both 'RMP off' and 'RMP on: n=3 0degree' profiles taken at the end of the inter-ELM period to n=5,10,15,20,25 modes. Since there is an overlap of pedestal widths in the coils on and off datasets, the stability analysis shown here is for cases with similar widths.**